\documentclass[12pt,thmsa]{article}

\begin{document}

\noindent {\LARGE The ''True Transformations Relativity'' Analysis of the
Michelson-Morley Experiment}{\Large \bigskip }

\noindent \textbf{Tomislav Ivezi\'{c}}

\noindent \textit{Ru{%
\mbox
 {\it{d}\hspace{-.15em}\rule[1.25ex]{.2em}{.04ex}\hspace{-.05em}}}er Bo\v
{s}kovi\'{c} Institute}

\noindent P.O.B. 180, 10002 Zagreb, Croatia

\noindent E-mail: ivezic@rudjer.irb.hr\textbf{\bigskip }

\noindent \rule[0.06in]{5.5in}{0.01in}

\noindent \textbf{Abstract}

\noindent In this paper we present an invariant formulation of special
relativity, i.e., the ''true transformations relativity.'' It deals either
with true tensor quantities (when no basis has been introduced) or
equivalently with coordinate-based geometric quantities comprising both
components and a basis (when some basis has been introduced). It is shown
that this invariant formulation, in which special relativity is understood
as the theory of a four-dimensional spacetime with the pseudo-Euclidean
geometry, completely explains the results of the Michelson-Morley
experiment. Two noncovariant approaches to the analysis of the
Michelson-Morley experiment are discussed; the coventional one in which only
the path lengths (optical or geometrical) are considered, and Driscoll's
approach (R.B. Driscoll, Phys. Essays \textbf{10, }394 (1997)), in which the
increment of phase is determined not only by the segment of geometric path
length, but also by the wavelength in that segment. Because these analyses
belong to the ''apparent transformations relativity,'' they do not agree
with the results of the Michelson-Morley experiment.

\noindent \rule[0.06in]{5.5in}{0.01in}

\newpage

\noindent {\large 1. INTRODUCTION}\textbf{\ }

Recently Driscoll$^{(1)}$ analyzed the Michelson-Morley$^{(2)}$ experiment
taking into account, in the calculation of the fringe shift, the Doppler
effect on wavelength in the frame in which the interferometer is moving. In
contrast to the traditional analysis the non-null fringe shift was found and
the author concluded: ''that the Maxwell-Einstein electromagnetic equations
and special relativity jointly are disproved, not confirmed, by the
Michelson-Morley experiment.'' In this paper we present an invariant
approach to special relativity (SR) with tensor quantities (and tensor
equations) to the analysis of the Michelson-Morley experiment and find a
null fringe shift in agreement with the experiment. We also show why the
calculation from$^{(1)}$ leads to the non-null result and why the
traditional analysis gives an apparent (not true) agreement with the
experiment.

In Sec. 2. we briefly discuss different approaches to SR. In the first
approach \emph{SR is formulated in terms of true tensor quantities and true
tensor equations}, which we call \emph{the ''true transformations (TT)
relativity.''} That approach is compared with \emph{the usual covariant
approach, which mainly deals with the basis components of tensors in a
specific, i.e., Einstein's coordinatization}$^{(3)}$ of the chosen inertial
frame of reference (IFR). The general discussion is illustrated in Sec. 2.1
by two examples: the spacetime length for a moving rod and the spacetime
length for a moving clock. The usual, i.e., Einstein's formulation of SR,
which is based on his two postulates, and which deals with the Lorentz
contraction and the dilatation of time, is also considered in Sec. 2. It is
shown that \emph{the Lorentz contraction and the dilatation of time are the
apparent transformations (AT).} (The notions of the TT and the AT are
introduced in Ref. 4.) \emph{Any approach to SR which uses the AT we call
the ''apparent transformations (AT) relativity.'' }Einstein's formulation of
SR obviously belongs to the ''AT relativity.'' The same two examples as
mentioned above are considered in the ''AT relativity'' in Sec. 2.2.

Then in Sec. 3.1 we discuss the nonrelativistic analysis and in Sec. 3.2 the
traditional analysis of the Michelson-Morley experiment. In Sec. 3.3 we
repeat in short Driscoll's calculation of the fringe shift in the
Michelson-Morley experiment. Both, the traditional analysis and Driscoll's
analysis of the Michelson-Morley experiment are shown to belong to the ''AT
relativity.'' In Sec. 4. we present the analysis of the Michelson-Morley
experiment in the ''TT relativity'' explicitly using two very different
synchronizations of distant clocks (they are explained in Sec. 2.) and we
find the null result in agreement with the experiment. It is important to
note that this result holds in all permissible coordinatizations since the
whole phase of a light wave, the true tensor, $\phi =k^{a}g_{ab}l^{b}$ (see
Eqs. (\ref{phase}) and (\ref{pha2})) is used in the calculation. In Sec. 4.1
we explain Driscoll's non-null fringe shift as a consequence of an ''AT
relativity'' calculation of the increment of phase. Driscoll's calculation
takes into account only a part $k^{0}l_{0}$ of the above mentioned whole
phase $\phi $ and considers that part in two relatively moving IFRs $S$ ($%
k^{0}l_{0}$) and $S^{\prime }$ ($k^{0^{\prime }}l_{0^{\prime }}$) but only
in the Einstein coordinatization (the $S$ frame is the interferometer rest
frame). Finally in Sec. 4.2 we explicitly show that the agreement with
experiment obtained in the traditional ''AT relativity'' calculation is
actually an apparent agreement. This calculation also deals with the part $%
k^{0}l_{0}$ of the whole phase $\phi $ and considers that part in IFRs $S$
and $S^{\prime }$ but again only in the Einstein coordinatization. In
contrast to Driscoll's calculation the traditional analysis considers the
contribution $k^{0}l_{0}$ in the interferometer rest frame $S,$ but in the $%
S^{\prime }$ frame, in which the interferometer is moving, it considers the
contribution $k^{0}l_{0^{\prime }}$; the $k^{0}$ factor is taken to be the
same in $S$ and $S^{\prime }$ frames. This fact caused an apparent agreement
of the traditional analysis with the results of the Michelson-Morley
experiment. Since only a part of the whole phase $\phi $ is considered both
results, Driscoll's and the traditional one, are synchronization, i.e.,
coordinatization, dependent results. Thus the agreement between the
traditional analysis and the experiment exists only when Einstein's
synchronization of distant clocks is used and not for another
synchronization. This is also proved in Sec. 4.2. \bigskip \medskip 

\noindent {\large 2. THE COMPARISON\ OF\ THE\ ''TT\ RELATIVITY'' WITH\ THE\
USUAL\ COVARIANT\ APPROACH\ AND\ WITH\ THE\ ''AT\ RELATIVITY''}

The above mentioned approaches to SR are partly discussed in Refs. 5-8.
Rohrlich,$^{(4)}$ and also Gamba,$^{(9)}$ emphasized the role of\ \emph{the
concept of sameness of a physical quantity for different observers.} This
concept determines the difference between the mentioned approaches and also
it determines what is to be understood as a relativistic theory. Our
invariant approach to SR, i.e., the ''TT relativity,'' and the concept of
sameness of a physical quantity for different observers in that approach,
differs not only from the ''AT relativity'' approach but also from the usual
covariant approach (including Refs. 4 and 9).

We first explain the difference between the ''TT relativity'' and the usual
covariant approach to SR. In the ''TT relativity'' SR is understood as the
theory of a four-dimensional (4D) spacetime with pseudo-Euclidean geometry.
All physical quantities (in the case when no basis has been introduced) are
described by \emph{true tensor fields}, that are defined on the 4D
spacetime, and that satisfy \emph{true tensor equations }representing
physical laws. When the coordinate system has been introduced the physical
quantities are mathematically represented by \emph{the coordinate-based
geometric quantities }(CBGQs) that satisfy \emph{the coordinate-based
geometric equations} (CBGEs). \emph{The CBGQs contain both the components
and the basis one-forms and vectors} of the chosen IFR. (Speaking in
mathematical language a tensor of type (k,l) is defined as a linear function
of k one-forms and l vectors (in old names, k covariant vectors and l
contravariant vectors) into the real numbers, see, e.g., Refs. 10-12. If a
coordinate system is chosen in some IFR then, in general, any tensor
quantity can be reconstructed from its components and from the basis vectors
and basis 1-forms of that frame, i.e., it can be written in a
coordinate-based geometric language, see, e.g., Ref. 12.) The symmetry
transformations for the metric $g_{ab}$, i.e., the isometries$^{(10)}$,
leave the pseudo-Euclidean geometry of 4D spacetime of SR unchanged. At the
same time they do not change the true tensor quantities, or equivalently the
CBGQs, in physical equations. Thus \emph{isometries are} what Rohrlich$%
^{(4)} $ calls \emph{the TT}. In the ''TT relativity'' different
coordinatizations of an IFR are allowed and they are all equivalent in the
description of physical phenomena. Particularly two very different
coordinatizations, the Einstein (''e'')$^{(3)}$ and ''radio'' (''r'')$%
^{(13)} $ coordinatization, will be briefly exposed and exploited in the
paper. In the ''e'' coordinatization the Einstein synchronization$^{(3)}$ of
distant clocks and cartesian space coordinates $x^{i}$ are used in the
chosen IFR. The main features of the ''r'' coordinatization will be given
below, see also.Refs. 13, 6 and 7. \emph{The CBGQs representing some 4D
physical quantity in different relatively moving IFRs, or in different
coordinatizations of the chosen IFR, are all mathematically equal since they
are connected by the TT (i.e., the isometries). }Thus they are really the
same quantity for different observers, or in different coordinatizations.
Hence it is appropriate to call the ''TT relativity'' approach (which deals
with the true tensors or with the CBGQs) as an invariant approach in
contrast to the usual covariant approach (which deals with the components of
tensors taken in the ''e'' coordinatization). We suppose that in the ''TT
relativity'' such 4D tensor quantities are well-defined not only
mathematically but also \emph{experimentally}, as measurable quantities with
real physical meaning. \emph{The complete and well-defined measurement from
the ''TT relativity'' viewpoint is such measurement in which all parts of
some 4D quantity are measured.}

However \emph{in the usual covariant approach one does not deal with the
true tensors, or equivalently with CBGQs, but with the basis components of
tensors (mainly in the ''e'' coordinatization) and with the equations of
physics written out in the component form. }Mathematically speaking the
concept of a tensor in the usual covariant approach is defined entirely in
terms of the transformation properties of its \emph{components} relative to
some coordinate system. The definitions of the same quantity in Refs. 4 and
9 also refer to such component form in the ''e'' coordinatization of tensor
quantities and tensor equations. It is true that the components of some
tensor refer to the same tensor quantity considered in two relatively moving
IFRs $S$ and $S^{\prime }$ and in the ''e'' coordinatization, but they
cannot be equal, since the bases are not included.

The third approach to SR uses the AT of some quantities. In contrast to the
TT the AT are not the transformations of spacetime tensors and they do not
refer to the same quantity. Thus they are not isometries and they \emph{%
refer exclusively to the component form of tensor quantities and in that
form they transform only some components of the whole tensor quantity.} In
fact, depending on the used AT, only a part of a 4D tensor quantity is
transformed by the AT. Such a part of a 4D quantity, when considered in
different IFRs (or in different coordinatizations of some IFR) corresponds
to different quantities in 4D spacetime. \emph{Some examples of\ the AT are:
the AT of the synchronously defined spatial length,}$^{(3)}$\emph{\ i.e.,
the Lorentz contraction}$^{(4-9)}$\emph{\ and the AT of the temporal
distance, i.e., the conventional dilatation of time} that is introduced in
Ref. 3 and considered in Refs. 7 and 8. \emph{The formulation of SR which
uses the AT we call the ''AT relativity.'' An example of such formulation is
Einstein's formulation of SR which is based on his two postulates and which
deals with all the mentioned AT.}

The differences between the ''TT relativity,'' the usual covariant approach
and the ''AT relativity'' will be examined considering some specific
examples. First the spacetime lengths, corresponding in ''3+1'' picture to a
moving rod and to a moving clock, will be considered in the ''TT
relativity.'' Furthermore the spatial and temporal distances for the same
examples will be examined in the ''AT relativity.'' The comparison with the
experiments on the Lorentz contraction and the time dilatation is performed
in Ref. 8 and it shows that all experiments can be qualitatively and
quantitatively explained by the ''TT relativity,'' while some experiments
cannot be adequately explained by the ''AT relativity.'' This will be also
shown below considering Michelson-Morley experiment.

Before doing this exploration we discuss the notation, different
coordinatizations and the connections between them. In this paper I use the
following convention with regard to indices. Repeated indices imply
summation. Latin indices $a,b,c,d,...$ are to be read according to the
abstract index notation, see Ref. 10 Sec. 2.4; they ''...should be viewed as
reminders of the number and type of variables the tensor acts on, \emph{not}
as basis components.'' They designate geometric objects in 4D spacetime.
Thus, e.g., $l_{AB}^{a}$ and $x_{A}^{a}$ (a distance 4-vector $%
l_{AB}^{a}=x_{B}^{a}-x_{A}^{a}$ between two events $A$ and $B$ whose
position 4-vectors are $x_{A}^{a}$ and $x_{B}^{a}$) are (1,0) tensors and
they are defined independently of any coordinate system. Greek indices run
from 0 to 3, while latin indices $i,j,k,l,...$ run from 1 to 3, and they
both designate the components of some geometric object in some coordinate
system, e.g., $x^{\mu }(x^{0},x^{i})$ and $x^{\mu ^{\prime }}(x^{0^{\prime
}},x^{i^{\prime }})$ are two coordinate representations of the position
4-vector $x^{a}$ in two different inertial coordinate systems $S$ and $%
S^{\prime }.$ Similarly the metric tensor $g_{ab}$ denotes a tensor of type
(0,2) (whose Riemann curvature tensor $R_{bcd}^{a}$ is everywhere vanishing;
the spacetime of SR is a flat spacetime, and this definition includes not
only the IFRs but also the accelerated frames of reference). This geometric
object $g_{ab}$ is represented in the component form in some IFR $S,$ and in
the ''e'' coordinatization, i.e., in the $\left\{ e_{\mu }\right\} $ basis,
by the $4\times 4$ diagonal matrix of components of $g_{ab}$, $g_{\mu \nu
,e}=diag(-1,1,1,1),$ and this is usually called the Minkowski metric tensor
(the subscript $^{\prime }e^{\prime }$ stands for the Einstein
coordinatization).

Different coordinatizations of some reference frame can be obtained using,
e.g., different synchronizations. On the other hand different
synchronizations are determined by the parameter $\varepsilon $ in the
relation $t_{2}=t_{1}+\varepsilon (t_{3}-t_{1})$, where $t_{1}$ and $t_{3}$
are the times of departure and arrival, respectively, of the light signal,
read by the clock at $A$, and $t_{2}$ is the time of reflection at $B$, read
by the clock at $B$, that has to be synchronized with the clock at $A$. In
Einstein's synchronization convention $\varepsilon =1/2.$ We can also choose
another coordinatization, the ''everyday'' or ''radio'' (''r'')
cordinatization,$^{(13)}$ which differs from the ''e'' coordinatization by
the different procedure for the synchronization of distant clocks. In the
''r'' synchronization $\varepsilon =0$ and thus, in contrast to the ''e''
synchronization, there is an absolute simultaneity. As explained in Ref. 13:
''For if we turn on the radio and set our clock by the standard announcement
''...at the sound of the last tone, it will be 12 o'clock'', then we have
synchronized our clock with the studio clock in a manner that corresponds to
taking $\varepsilon =0$ in $t_{2}=t_{1}+\varepsilon (t_{3}-t_{1}).$'' The
''r'' synchronization is an assymetric synchronization which leads to an
assymetry in the coordinate, one-way, speed of light.$^{(13)}$ However from
the physical point of view the ''r'' coordinatization is completely
equivalent to the ''e'' coordinatization. This also holds for all other
permissible coordinatizations. Such situation really happens in the ''TT
relativity.'' As explained above the ''TT relativity'' deals with true
tensors and the true tensor equations (when no basis has been chosen)\emph{,}
or equivalently (when the coordinate basis has been introduced) with the
CBGQs and CBGEs. Thus the ''TT relativity'' deals on the same footing with
all possible coordinatizations of the chosen reference frame. As a
consequence \emph{the second Einstein postulate referred to the constancy of
the coordinate velocity of light, in general, does not hold in the ''TT
relativity.''} Namely, only in Einstein's coordinatization the coordinate,
one-way, speed of light is isotropic and constant.

In the following we shall also need the expression for the covariant 4D
Lorentz transformations $L^{a}{}_{b}$, which is independent of the chosen
synchronization, i.e., coordinatization of reference frames, see the works.$%
^{(14,6,7)}$ It is 
\begin{equation}
L^{a}{}_{b}\equiv
L^{a}{}_{b}(v)=g^{a}{}_{b}-((2u^{a}v_{b})/c^{2})+(u^{a}+v^{a})(u_{b}+v_{b})/c^{2}(1+\gamma ),
\label{fah}
\end{equation}
where $u^{a}$ is the proper velocity 4-vector of a frame $S$ with respect to
itself (only $u^{0}\neq 0$, see also Ref. 13) $u^{a}=cn^{a},$ $n^{a}$ is the
unit 4-vector along the $x^{0}$ axis of the frame $S,$ and $v^{a}$ is the
proper velocity 4-vector of $S^{\prime }$ relative to $S.$ Further $u\cdot
v=u^{a}v_{a}$ and $\gamma =-u\cdot v/c^{2}.$ When we use the ''e''
coordinatization then $L^{a}{}_{b}$ is represented by $L^{\mu }{}_{\nu ,e},$
the usual expression for pure Lorentz transformation, but with $v_{e}^{i}$
(the proper velocity 4-vector $v_{e}^{\mu }$ is $v_{e}^{\mu }\equiv
dx_{e}^{\mu }/d\tau =(\gamma _{e}c,\gamma _{e}v_{e}^{i}),$ $d\tau \equiv
dt_{e}/\gamma _{e}$ is the scalar proper-time, and $\gamma _{e}\equiv
(1-v_{e}^{2}/c^{2})^{1/2}$) replacing the components of the ordinary
velocity 3-vector $\mathbf{V.}$ Obviously, in the usual form, the Lorentz
transformations connect two coordinate representations, basis components (in
the ''e'' coordinatization) $x_{e}^{\mu },$ $x_{e}^{\mu ^{\prime }}$ of a
given event; $x_{e}^{\mu },$ $x_{e}^{\mu ^{\prime }}$ refer to two
relatively moving IFRs (with the Minkowski metric tensors) $S$ and $%
S^{\prime },$ 
\begin{eqnarray}
x_{e}^{\mu ^{\prime }} &=&L^{\mu ^{\prime }}{}_{\nu ,e}x_{e}^{\nu },\qquad
\,L^{0^{\prime }}{}_{0,e}=\gamma _{e},\quad L^{0^{\prime
}}{}_{i,e}=L^{i^{\prime }}{}_{0,e}=-\gamma _{e}v_{e}^{i}/c,  \nonumber \\
L^{i^{\prime }}{}_{j,e} &=&\delta _{j}^{i}+(\gamma
_{e}-1)v_{e}^{i}v_{je}/v_{e}^{2}.  \label{lorus}
\end{eqnarray}
Since $g_{\mu \nu ,e}$ is a diagonal matrix the space $x_{e}^{i}$ and time $%
t_{e}$ $(x_{e}^{0}\equiv ct_{e})$ parts of $x_{e}^{\mu }$ do have their
usual meaning.

The invariant spacetime length (the Lorentz scalar) between two points
(events) in 4D spacetime is defined as 
\begin{equation}
l=(g_{ab}l^{a}l^{b})^{1/2},  \label{elspat}
\end{equation}
where $l^{a}(l^{b})$ is the distance 4-vector between two events $A$ and $B$%
, $l^{a}=l_{AB}^{a}=x_{B}^{a}-x_{A}^{a}$. In the ''e'' coordinatization the
geometrical quantity $l^{2}$ can be written in terms of its representation $%
l_{e}^{2},$ with the separated spatial and temporal parts, $%
l^{2}=l_{e}^{2}=(l_{e}^{i}l_{ie})-(l_{e}^{0})^{2}$. Such separation remains
valid in other inertial coordinate systems with the Minkowski metric tensor,
and in $S^{\prime }$ one finds $l^{2}=l_{e}^{\prime 2}=(l_{e}^{i^{\prime
}}l_{i^{\prime }e})-(l_{e}^{0^{\prime }})^{2},$ where $l_{e}^{\mu ^{\prime
}} $ in $S^{\prime }$ is connected with $l_{e}^{\mu }$ in $S$ by the Lorentz
transformation $L^{\mu ^{\prime }}{}_{\nu ,e}$ (\ref{lorus}).

This is not so in the ''r'' cordinatization. In order to explain this
statement we now expose the ''r'' cordinatization in more detail. The basis
vectors in the ''r'' cordinatization are constructed as in Refs. 13, 6 and
7. The temporal basis vector $e_{0}$ is the unit vector directed along the
world line of the clock at the origin. The spatial basis vectors by
definition connect \emph{simultaneous} events, the event ''clock at rest at
the origin reads 0 time'' with the event ''clock at rest at unit distance
from the origin reads 0 time,'' and thus they are synchronization-dependent.
The spatial basis vector $e_{i}$ connects two above mentioned simultaneous
events when Einstein's synchronization ($\varepsilon =1/2$) of distant
clocks is used. The temporal basis vector $r_{0}$ is the same as $e_{0}.$
The spatial basis vector $r_{i}$ connects two above mentioned simultaneous
events when ''radio'' clock synchronization ($\varepsilon =0$) of distant
clocks is used. The spatial basis vectors, e.g., $r_{1},r_{1^{\prime
}}\,,r_{1^{\prime \prime }}..$ are parallel and directed along an
(observer-independent) light line. Hence, two events that are ''everyday''
(''r'') simultaneous in $S$ are also ''r'' simultaneous for all other IFRs.

The connection between the basis vectors in the ''r'' and ''e''
coordinatizations is given$^{(13,6,7)}$ as 
\[
r_{0}=e_{0},\;r_{i}=e_{0}+e_{i}. 
\]
The geometry of the spacetime is generally defined by the metric tensor $%
g_{ab},$ which can be expand in a coordinate basis in terms of its
components as $g_{ab}=g_{\mu \nu }dx^{\mu }\otimes dx^{\nu },$ and where $%
dx^{\mu }\otimes dx^{\nu }$ is an outer product of the basis 1-forms. The
metric tensor $g_{ab}$ becomes $g_{ab}=g_{\mu \nu ,r}dx_{r}^{\mu }\otimes
dx_{r}^{\nu }$ in the coordinate-based geometric language and in the ''r''
coordinatization, where the basis components of the metric tensor are 
\[
g_{00,r}=g_{0i,r}=g_{i0,r}=g_{ij,r}(i\neq j)=-1,\ g_{ii,r}=0. 
\]
$dx_{r}^{\mu },$ $dx_{r}^{\nu }$ are the basis 1-forms in the ''r''
coordinatization and in $S,$ and $dx_{r}^{\mu }\otimes dx_{r}^{\nu }$ is an
outer product of the basis 1-forms, i.e., it is the basis for (0,2) tensors
(the subscript $^{\prime }r^{\prime }$ stands for the ''r''
coordinatization).

The transformation matrix$^{(6,7)}$ $T_{\;\nu ,r}^{\mu }$ transforms the
''e'' coordinatization to the ''r'' coordinatization. The elements that are
different from zero are 
\[
T_{\;\mu ,r}^{\mu }=-T_{\;i,r}^{0}=1.
\]
For the sake of completeness we also quote the Lorentz transformation $%
L^{\mu ^{\prime }}\,_{\nu ,r}$ in the ''r'' coordinatization. It can be
easily found from $L^{a}{}_{b}$ (\ref{fah}) and the known $g_{\mu \nu ,r},$
and the elements that are different from zero are 
\begin{eqnarray}
x_{r}^{\prime \mu } &=&L^{\mu ^{\prime }}{}_{\nu ,r}x_{r}^{\nu },\quad
L^{0^{\prime }}{}_{0,r}=K,\quad L^{0^{\prime }}{}_{2,r}=L^{0^{\prime
}}{}_{3,r}=K-1,  \nonumber \\
L^{1^{\prime }}{}_{0,r} &=&L^{1^{\prime }}{}_{2,r}=L^{1^{\prime
}}{}_{3,r}=(-\beta _{r}/K),L^{1^{\prime }}{}_{1,r}=1/K,  \nonumber \\
L^{2^{\prime }}{}_{2,r} &=&L^{3^{\prime }}{}_{3,r}=1,  \label{elr}
\end{eqnarray}
$L^{1^{\prime }}{}_{0,r}=L^{1^{\prime }}{}_{2,r}=L^{1^{\prime
}}{}_{3,r}=(-\beta _{r}/K),L^{1^{\prime }}{}_{1,r}=1/K,$ $L^{2^{\prime
}}{}_{2,r}=L^{3^{\prime }}{}_{3,r}=1,$where $K=(1+2\beta _{r})^{1/2},$ and $%
\beta _{r}=dx_{r}^{1}/dx_{r}^{0}$ is the velocity of the frame $S^{\prime }$
as measured by the frame $S$, $\beta _{r}=\beta _{e}/(1-\beta _{e})$ and it
ranges as $-1/2\prec \beta _{r}\prec \infty .$ Since $g_{\mu \nu ,r},$ in
contrast to $g_{\mu \nu ,e},$ is not a diagonal matrix, then in the
spacetime length $l,$ i.e., $l^{2},$ the spatial and temporal parts are not
separated. Expressing $l_{r}^{\mu }$ in terms of $l_{e}^{\mu }$ one finds
that $l^{2}=l_{r}^{2}=l_{e}^{2},$ as it must be. It can be easily proved$%
^{(13)}$ that the ''r'' synchronization is an assymetric synchronization
which leads to an assymetry in the measured ''one-way'' velocity of light
(for one direction $c_{r}^{+}=\infty $ whereas in the opposite direction $%
c_{r}^{-}=-c/2$). The round trip velocity, however, does not depend on the
chosen synchronization procedure, and it is $\equiv c.$ Although in the
''e'' coordinatization the space and time components of the position
4-vector do have their usual meaning, i.e., as in the prerelativistic
physics, and in $l_{e}^{2}$ the spatial and temporal parts are separated, it
does not mean that the ''e'' coordinatization does have some advantage
relative to other coordinatizations and that the quantities in the ''e''
coordinatization are more physical.

A symmetry transformation for the metric $g_{ab}$ is called an isometry and
it does not change $g_{ab}$; if we denote an isometry as $\Phi ^{*}$ then $%
(\Phi ^{*}g)_{ab}=g_{ab}.$ Thus an isometry leaves the pseudo-Euclidean
geometry of 4D spacetime of SR unchanged. An example of isometry is the
covariant 4D Lorentz transformation $L^{a}{}_{b}$ (\ref{fah}). In our
terminology \emph{the TT are nothing else but - the isometries.} When the
coordinate basis is introduced then, for example, the isometry $L^{a}{}_{b}$
(\ref{fah}) will be expressed as the isometries, the coordinate Lorentz
transformation $L^{\mu ^{\prime }}{}_{\nu ,e}$ (\ref{lorus}) in the ''e''
coordinatization, or as $L^{\mu ^{\prime }}\,_{\nu ,r}$ (\ref{elr}) in the
''r'' coordinatization. In our treatment mainly the coordinate-based
geometric form will be used for tensors representing physical quantities and
for tensor equations representing physical laws. The basis components of the
CBGQs will be transformed, e.g., by $L^{\mu ^{\prime }}{}_{\nu ,e}$ while
the basis vectors $e_{\mu }$ by the inverse transformation $(L^{\mu ^{\prime
}}{}_{\nu ,e})^{-1}=L^{\mu }{}_{\nu ^{\prime },e}.$

The above consideration enable us to better explain the difference in the
concept of \emph{sameness} of a physical quantity for the ''TT relativity''
approach and the usual covariant approach. We consider a simple example the
distance 4-vector (the $(1,0)$ tensor) $l_{AB}^{a}=x_{B}^{a}-x_{A}^{a}$
between two events $A$ and $B$ (with the position 4-vectors $x_{A}^{a}$ and $%
x_{B}^{a}$). It can be equivalently represented in the coordinate-based
geometric language in different bases, $\left\{ e_{\mu }\right\} $ and $%
\left\{ r_{\mu }\right\} $ in an IFR $S,$ and $\left\{ e_{\mu ^{\prime
}}\right\} $ and $\left\{ r_{\mu ^{\prime }}\right\} $ in a relatively
moving IFR $S^{\prime },$ as $l_{AB}^{a}=l_{e}^{\mu }e_{\mu }=l_{r}^{\mu
}r_{\mu }=l_{e}^{\mu ^{\prime }}e_{\mu ^{\prime }}=l_{r}^{\mu ^{\prime
}}r_{\mu ^{\prime }},$ where, e.g., $e_{\mu }$ are the basis 4-vectors, $%
e_{0}=(1,0,0,0)$ and so on, and $l_{e}^{\mu }$ are the basis components when
the ''e'' coordinatization is chosen in some IFR $S.$ The decompositions $%
l_{e}^{\mu }e_{\mu }$ and $l_{r}^{\mu }r_{\mu }$ (in an IFR $S,$ and in the
''e'' and ''r'' coordinatizations respectively) and $l_{e}^{\mu ^{\prime
}}e_{\mu ^{\prime }}$ and $l_{r}^{\mu ^{\prime }}r_{\mu ^{\prime }}$ (in a
relatively moving IFR $S^{\prime }$, and in the ''e'' and ''r''
coordinatizations respectively) of the true tensor $l_{AB}^{a}$ are all
mathematically \emph{equal} quantities. Thus they are really the same
quantity considered in different relatively moving IFRs and in different
coordinatizations. This is the treatment of the distance 4-vector in the
''TT relativity.'' On the other hand the usual covariant approach does not
consider the whole tensor quantity, the distance 4-vector $l_{AB}^{a},$ but
only the basis components, mainly $l_{e}^{\mu }$ and $l_{e}^{\mu ^{\prime }}$
in the ''e'' coordinatization (or$^{(6-8)}$ $l_{r}^{\mu }$ and $l_{r}^{\mu
^{\prime }}$ in the ''r'' coordinatization). Note that, in contrast to the
above equalities for the CBGQs, the sets of components, e.g., $l_{e}^{\mu }$
and $l_{e}^{\mu ^{\prime }},$ taken alone, are not equal, $l_{e}^{\mu }\neq
l_{e}^{\mu ^{\prime }},$ and thus they are not the same quantity from the
''TT relativity'' viewpoint. From the mathematical point of view the
components of, e.g., a $(1,0)$ tensor are its values (real numbers) when the
basis one-form, for example, $e^{\alpha },$ is its argument (see, e.g., Ref.
11). Thus, for example, $l_{AB}^{a}(e^{\alpha })=l_{e}^{\mu }e_{\mu
}(e^{\alpha })=l_{e}^{\alpha }$ (where $e^{\alpha }$ is the basis one-form
in an IFR $S$ and in the ''e'' coordinatization), while $l_{AB}^{a}(e^{%
\alpha ^{\prime }})=l_{e}^{\mu ^{\prime }}e_{\mu ^{\prime }}(e^{\alpha
^{\prime }})=l_{e}^{\alpha ^{\prime }}$ (where $e^{\alpha ^{\prime }}$ is
the basis one-form in $S^{\prime }$ and in the ''e'' coordinatization).
Obviously $l_{e}^{\alpha }$ and $l_{e}^{\alpha ^{\prime }}$ are not the same
real numbers since the basis one-forms $e^{\alpha }$ and $e^{\alpha ^{\prime
}}$ are different bases.\bigskip

\noindent \textbf{2.1 The\ TT of\ the\ Spacetime\ Length }

In order to explore the difference between the ''TT relativity'' and the
''AT relativity'' we consider the spacetime length for a moving rod and then
for a moving clock. The same examples will be also examined in the ''AT
relativity.''. Let us take, for simplicity, to work in 2D spacetime. Then we
also take a particular choice for the 4-vector $l_{AB}^{a}.$ In the usual
''3+1'' picture it corresponds to an object, a rod, that is at rest in an
IFR $S$ and situated along the common $x_{e}^{1},x_{e}^{1^{\prime }}-$ axes; 
$L_{0}$ is its rest length. The decomposition of the geometric quantity $%
l_{AB}^{a}$ in the ''e'' coordinatization and in $S$ is $%
l_{AB}^{a}=l_{e}^{0}e_{0}+l_{e}^{1}e_{1}=0e_{0}+L_{0}e_{1},$ while in $%
S^{\prime },$ where the rod is moving, it becomes $l_{AB}^{a}=-\beta
_{e}\gamma _{e}L_{0}e_{0^{\prime }}+\gamma _{e}L_{0}e_{1^{\prime }},$ and,
as explained above, it holds that 
\begin{equation}
l_{AB}^{a}=0e_{0}+L_{0}e_{1}=-\beta _{e}\gamma _{e}L_{0}e_{0^{\prime
}}+\gamma _{e}L_{0}e_{1^{\prime }}.  \label{trucon}
\end{equation}
$l_{AB}^{a}$ is a tensor of type (1,0) and in (\ref{trucon}) it is written
in the coordinate-based geometric language in terms of basis vectors $e_{0},$
$e_{1},$ ($e_{0^{\prime }},$ $e_{1^{\prime }}$) and the basis components $%
l_{e}^{\mu }$ ($l_{e}^{\mu ^{\prime }}$) of a specific IFR.

We note once again that in the ''TT relativity'' the basis components $%
l_{e}^{\mu }$\ in $S$\ and $l_{e}^{\mu ^{\prime }}$\ in $S^{\prime },$\ when
taken alone, do not represent the same 4D quantity. Only the geometric
quantity $l_{AB}^{a},$\ i.e., the CBGQs $l_{e}^{\mu }e_{\mu }=l_{e}^{\mu
^{\prime }}e_{\mu ^{\prime }}$ comprising both, components and a basis, is
the same 4D quantity for different relatively moving IFRs; Ref. 11:
''....the components tell only part of the story. The basis contains the
rest of information.'' Of course we could equivalently work in another
coordinatization, e.g., the ''r'' coordinatization, as shown in Refs. 6 and
7. Then we would find that $l_{AB}^{a}=l_{e}^{\mu }e_{\mu }=l_{r}^{\mu
}r_{\mu }=l_{e}^{\mu ^{\prime }}e_{\mu ^{\prime }}=l_{r}^{\mu ^{\prime
}}r_{\mu ^{\prime }},$ where $r_{\mu }$ and $r_{\mu ^{\prime }}$ are the
basis 4-vectors, and $l_{r}^{\mu }$ and $l_{r}^{\mu ^{\prime }}$ are the
basis components in the ''r'' coordinatization, and in $S$ and $S^{\prime }$
respectively. (The expressions for $l_{r}^{\mu }$ and $l_{r}^{\mu ^{\prime
}} $ can be easily found from the known transformation matrix $T_{\;\nu
,r}^{\mu }$.) We see from (\ref{trucon}) that in the ''e'' coordinatization,
which is commonly used in the ''AT relativity,'' there is a dilatation of
the spatial part $l_{e}^{1^{\prime }}=\gamma _{e}L_{0}$ with respect to $%
l_{e}^{1}=L_{0}$ and not the Lorentz contraction as predicted in the ''AT
relativity.'' Hovewer it is clear from the above discussion that comparison
of only spatial parts of the components of the distance 4-vector $l_{AB}^{a}$
in $S$ and $S^{\prime }$ is physically meaningless in the ''TT relativity.''
When only some components of the whole tensor quantity are taken alone,
then, in the ''TT relativity,'' they do not represent some definite physical
quantity in the 4D spacetime. Also we remark that always the \emph{whole
tensor quantity} $l_{AB}^{a}$\ comprising components and a basis is
transformed by the Lorentz transformation from $S$\ to $S^{\prime }.$ Note
that if $l_{e}^{0}=0$ then $l_{e}^{\mu ^{\prime }}$ in any other IFR $%
S^{\prime }$ will contain the time component $l_{e}^{0^{\prime }}\neq 0.$
The spacetime length for the considered case is frame and coordinatization
independent quantity $l=(l_{e,r}^{\mu }l_{\mu e,r})^{1/2}=(l_{e,r}^{\mu
^{\prime }}l_{\mu ^{\prime }e,r})^{1/2}=L_{0}.$ In the ''e''
coordinatization and in $S,$ the rest frame of the rod, where the temporal
part of $l_{e}^{\mu }$ is $l_{e}^{0}=0,$ the spacetime length $l$ is a
measure of the spatial distance, i.e., of the rest spatial length of the
rod, as in the prerelativistic physics.

In a similar manner we can choose another particular choice for the distance
4-vector $l_{AB}^{a},$ which will correspond to the well-known ''muon
experiment,'' and which is interpreted in the ''AT relativity'' in terms of
the time dilatation. First we consider this example in the ''TT
relativity.'' The distance 4-vector $l_{AB}^{a}$ will be examined in two
relatively moving IFRs $S$ and $S^{\prime }$, i.e., in the $\left\{ e_{_{\mu
}}\right\} $ and $\left\{ e_{\mu ^{\prime }}\right\} $ bases. The $S$ frame
is chosen to be the rest frame of the muon. Two events are considered; the
event $A$ represents the creation of the muon and the event $B$ represents
its decay after the lifetime $\tau _{0}$ in $S.$ The position 4-vectors of
the events $A$ and $B$ in $S$ are taken to be on the world line of a
standard clock that is at rest in the origin of $S.$ The distance 4-vector $%
l_{AB}^{a}=x_{B}^{a}-x_{A}^{a}$ that connects the events $A$ and $B$ is
directed along the $e_{0}$ basis vector from the event $A$ toward the event $%
B.$ This geometric quantity can be written in the coordinate-based geometric
language. Thus it can be decomposed in two bases $\left\{ e_{\mu }\right\} $
and $\left\{ e_{\mu ^{\prime }}\right\} $ as 
\begin{equation}
l_{AB}^{a}=c\tau _{0}e_{0}+0e_{1}=\gamma c\tau _{0}e_{0^{\prime }}+\beta
_{e}\gamma _{e}c\tau _{0}e_{1^{\prime }}.  \label{comu}
\end{equation}
Similarly we can easily find$^{(6,7)}$ the decompositions of $l_{AB}^{a}$ in
the ''r'' coordinatization. We again see that these decompositions,
containing both the basis components and the basis vectors, are the same
geometric quantity $l_{AB}^{a}.$ $l_{AB}^{a}$ does have only temporal parts
in $S$, while in the $\left\{ e_{\mu ^{\prime }}\right\} $ basis $l_{AB}^{a}$
contains not only the temporal part but also the spatial part. It is visible
from (\ref{comu}) that the comparison of only temporal parts of the
representations of the distance 4-vector is physically meaningless in the
''TT relativity.'' The spacetime length $l$ is always a well-defined
quantity in the ''TT relativity'' and for this example it is $l=(l_{e}^{\mu
}l_{\mu e})^{1/2}=(l_{e}^{\mu ^{\prime }}l_{\mu ^{\prime
}e})^{1/2}=(l_{r}^{\mu }l_{\mu r})^{1/2}=(l_{r}^{\mu ^{\prime }}l_{\mu
^{\prime }r})^{1/2}=(-c^{2}\tau _{0}^{2})^{1/2}$. Since in $S$ the spatial
parts $l_{e,r}^{1}$ of $l_{e,r}^{\mu }$ are zero the spacetime length $l$ in 
$S$ is a measure of the temporal distance, as in the prerelativistic
physics; one defines that $c^{2}\tau _{0}^{2}=-l_{e}^{\mu }l_{\mu
e}=-l_{r}^{\mu }l_{\mu r}.\bigskip $

\noindent \textbf{2.2 The\ AT of the\ Spatial\ and\ Temporal\ Distances }

In order to better explain the difference between the TT and the AT we now
consider the same two examples as above but from the point of view of the
conventional, i.e., Einstein's$^{(3)}$ interpretations of \emph{the} \emph{%
spatial length} of the moving rod and \emph{the temporal distance} for the
moving clock.

The synchronous definition of \emph{the spatial length}, introduced by
Einstein,$^{(3)}$ defines length as \emph{the spatial distance} between two
spatial points on the (moving) object measured by simultaneity in the rest
frame of the observer. One can see that the concept of sameness of a
physical quantity is quite different in the ''AT relativity'' but in the
''TT relativity.'' Thus for the Einstein definition of \emph{the spatial
length} one considers only \emph{the component} $l_{e}^{1}=L_{0}$ of $%
l_{e}^{\mu }e_{\mu }$ (when $l_{e}^{0}$ is taken $=0,$ i.e., the spatial
ends of the rod at rest in $S$ are taken simultaneously at $t=0$) and
performs the Lorentz transformation $L^{\mu }{}_{\nu ,e}$ of the basis
components $l_{e}^{\mu }$ (but not of the basis itself) from $S^{\prime }$
to $S,$ which yields 
\begin{eqnarray}
l_{e}^{0} &=&\gamma _{e}l_{e}^{0^{\prime }}+\gamma _{e}\beta
_{e}l_{e}^{1^{\prime }}  \nonumber \\
l_{e}^{1} &=&\gamma _{e}l_{e}^{1^{\prime }}+\gamma _{e}\beta
_{e}l_{e}^{0^{\prime }}.  \label{elcon}
\end{eqnarray}
Then one retains only the transformation of the spatial component $l_{e}^{1}$
(the second equation in (\ref{elcon})) \emph{neglecting completely the
transformation of the temporal part} $l_{e}^{0}$ (the first equation in (\ref
{elcon})). Furthermore in the transformation for $l_{e}^{1}$ one takes that
the temporal part in $S^{\prime }$ $l_{e}^{0^{\prime }}=0,$ ( i.e., the
spatial ends of the rod moving in $S^{\prime }$ are taken simultaneously at
some \emph{arbitrary }$t^{\prime }=b$). The quantity obtained in such a way
will be denoted as $L_{e}^{1^{\prime }}$ (it is not equal to $%
l_{e}^{1^{\prime }}$ appearing in the transformation equations (\ref{elcon}%
)) This quantity $L_{e}^{1^{\prime }}$ defines in the ''AT relativity'' 
\emph{the synchronously determined spatial length }of the moving rod in $%
S^{\prime }$. The mentioned procedure gives $l_{e}^{1}=\gamma
_{e}L_{e}^{1^{\prime }},$ that is, the famous formula for the Lorentz
contraction, 
\begin{equation}
L_{e}^{1^{\prime }}=l_{e}^{1}/\gamma _{e}=L_{0}/\gamma _{e},  \label{apcon}
\end{equation}
This quantity, $L_{e}^{1^{\prime }}=L_{0}/\gamma _{e},$ is the usual Lorentz
contracted \emph{spatial length}$,$ and the quantities $L_{0}$ and $%
L_{e}^{1^{\prime }}=L_{0}/\gamma _{e}$ are considered in the ''AT
relativity'' to be \emph{the same quantity} for observers in $S$ and $%
S^{\prime }$. The comparison with the relation (\ref{trucon}) clearly shows
that the quantities $L_{0}$\emph{\ and }$L_{e}^{1^{\prime }}=L_{0}/\gamma
_{e}$\emph{\ are two different and independent quantities in 4D spacetime.}
Thus, \emph{in the ''TT relativity'' the same quantity for different
observers is the tensor quantity, the 4-vector }$l_{AB}^{a}=l_{e}^{\mu
}e_{\mu }=l_{e}^{\mu ^{\prime }}e_{\mu ^{\prime }}=l_{r}^{\mu }r_{\mu
}=l_{r}^{\mu ^{\prime }}r_{\mu ^{\prime }};$\emph{\ only one quantity in 4D
spacetime.} However \emph{in the ''AT relativity'' different quantities in
4D spacetime, the spatiall distances }$l_{e}^{1}=L_{0}\emph{\ }$\emph{and }$%
L_{e}^{1^{\prime }}$\emph{\ (or similarly in the ''r'' coordinatization)\
are considered as the same quantity for different observers.} The relation
for the Lorentz ''contraction'' of the moving rod in the ''r''
coordinatization can be easily obtained performing the same procedure as in
the ''e'' coordinatization, and it is 
\begin{equation}
L_{r}^{1^{\prime }}=L_{0}/K=(1+2\beta _{r})^{-1/2}L_{0},  \label{aper}
\end{equation}
see also Refs. 6 and 7. We see from (\ref{aper}) that there is a length
''dilatation'' $\infty \succ L_{r}^{1^{\prime }}\succ L_{0}$ for $-1/2\prec
\beta _{r}\prec 0$ and the standard length ''contraction'' $L_{0}\succ
L_{r}^{1^{\prime }}\succ 0$ for positive $\beta _{r},$ which clearly shows
that the Lorentz contraction is not physically correctly defined
transformation. \emph{Thus the Lorentz contraction is the transformation
that connects different quantities (in 4D spacetime) in }$S$\emph{\ and }$%
S^{\prime },$\emph{\ or in different coordinatizations, which implies that
it is - an AT. }

The same example of the ''muon decay'' will be now considered in the ''AT
relativity'' (see also$^{(7)}$). In the ''e'' coordinatization the events $A$
and $B$ are again on the world line of a muon that is at rest in $S.$ We
shall see once again that the concept of sameness of a physical quantity is
quite different in the ''AT relativity.'' Thus for this example one compares 
\emph{the basis component} $l_{e}^{0}=c\tau _{0}$ of $l_{e}^{\mu }e_{\mu }$
with the quantity, which is obtained from \emph{the basis component} $%
l_{e}^{0^{\prime }}$ in the following manner; first one performs the Lorentz
transformation of the basis components $l_{e}^{\mu }$ (but not of the basis
itself) from the muon rest frame $S$ to the frame $S^{\prime }$ in which the
muon is moving. This procedure yields 
\begin{eqnarray}
l_{e}^{0^{\prime }} &=&\gamma _{e}l_{e}^{0}-\gamma _{e}\beta _{e}l_{e}^{1} 
\nonumber \\
l_{e}^{1^{\prime }} &=&\gamma _{e}l_{e}^{1}-\gamma _{e}\beta _{e}l_{e}^{0}.
\label{eltime}
\end{eqnarray}
Similarly as in the Lorentz contraction \emph{one now forgets the
transformation of the spatial part} $l_{e}^{1^{\prime }}$ (the second
equation in (\ref{eltime})) and considers only the transformation of the
temporal part $l_{e}^{0^{\prime }}$ (the first equation in (\ref{eltime})).
This is, of course, an incorrect step from the ''TT relativity'' viewpoint.
Then taking that $l_{e}^{1}=0$ (i.e., that $x_{Be}^{1}=x_{Ae}^{1}$) in the
equation for $l_{e}^{0^{\prime }}$ (the first equation in (\ref{eltime}))
one finds the new quantity which will be denoted as $L_{e}^{0^{\prime }}$
(it is not the same as $l_{e}^{0^{\prime }}$ appearing in the transformation
equations (\ref{eltime})). The temporal distance $l_{e}^{0}$ defines in the
''AT relativity,'' and in the ''e'' basis, the muon lifetime at rest, while $%
L_{e}^{0^{\prime }}$ is considered in the ''AT relativity,'' and in the
''e'' coordinatization, to define the lifetime of the moving muon in $%
S^{\prime }.$ The relation connecting $L_{e}^{0^{\prime }}$ with $l_{e}^{0},$
which is obtained by the above procedure, is then the well-known relation
for the time dilatation, 
\begin{equation}
L_{e}^{0^{\prime }}/c=t_{e}^{\prime }=\gamma _{e}l_{e}^{0}/c=\tau
_{0}(1-\beta _{e}^{2})^{-1/2}.  \label{tidil}
\end{equation}
By the same procedure we can find$^{(7)}$ the relation for the time
''dilatation'' in the ''r'' coordinatization 
\begin{equation}
L_{r}^{0^{\prime }}=Kl_{r}^{0}=(1+2\beta _{r})^{1/2}c\tau _{0}.
\label{tider}
\end{equation}
This relation shows that the new quantity $L_{r}^{0^{\prime }},$ which
defines in the ''AT relativity'' the temporal separation in $S^{\prime },$
where the clock is moving, is smaller - time ''contraction'' - but the
temporal separation $l_{r}^{0}=c\tau _{0}$ in $S,$ where the clock is at
rest, for $-1/2\prec \beta _{r}\prec 0,$ and it is larger - time
''dilatation'' - for $0\prec \beta _{r}\prec \infty $. From this
consideration we conclude that \emph{in the ''TT relativity'' the same
quantity for different observers is the tensor quantity, the 4-vector }$%
l_{AB}^{a}=l_{e}^{\mu }e_{\mu }=l_{e}^{\mu ^{\prime }}e_{\mu ^{\prime
}}=l_{r}^{\mu }r_{\mu }=l_{r}^{\mu ^{\prime }}r_{\mu ^{\prime }};$\emph{\
only one quantity in 4D spacetime.} However \emph{in the ''AT relativity''
different quantities in 4D spacetime, the temporal distances }$l_{e}^{0},$%
\emph{\ }$L_{e}^{0^{\prime }},$\emph{\ }$l_{r}^{0},$\emph{\ }$%
L_{r}^{0^{\prime }}$ \emph{are considered as the same quantity for different
observers. This shows that the time ''dilatation'' is the transformation
connecting different quantities (in 4D spacetime) in }$S$\emph{\ and }$%
S^{\prime },$\emph{\ or in different coordinatizations, which implies that
it is - an AT. }

We can compare the obtained results for the determination of the spacetime
length in the ''TT relativity'' and the determination of the spatial and
temporal distances in the ''AT\ relativity'' with the existing experiments.
This comparison is presented in Ref. 8. It is shown there that the ''TT
relativity'' results agree with all experiments that are complete from the
''TT relativity'' viewpoint, i.e., in which all parts of the considered
tensor quantity are measured in the experiment. However the ''AT
relativity'' results agree only with some of examined experiments.

The difference between the ''AT\ relativity'' and the ''TT relativity'' will
be now examined considering the famous Michelson-Morley$^{(2)}$
experiment.\bigskip \medskip

\noindent {\large 3. THE MICHELSON-MORLEY EXPERIMENT}\textbf{\ }

In the Michelson-Morley$^{(2)}$ experiment two light beams emitted by one
source are sent, by half-silvered mirror $O$, in orthogonal directions.
These partial beams of light traverse the two equal (of the length $L$) and
perpendicular arms $OM_{1}$ (perpendicular to the motion) and $OM_{2}$ (in
the line of motion) of Michelson's interferometer. The behaviour of the
interference fringes produced on bringing together these two beams after
reflection on the mirrors $M_{1}$ and $M_{2}$ is examined. The experiment
consists of looking for a shift of the intereference fringes as the
apparatus is rotated. The expected maximum shift in the number of fringes
(the measured quantity) on a $90^{0}$\ rotation is 
\begin{equation}
\bigtriangleup N=\bigtriangleup (\phi _{2}-\phi _{1})/2\pi ,  \label{delfi}
\end{equation}
where $\bigtriangleup (\phi _{2}-\phi _{1})$ is the change in the phase
difference when the interferometer is rotated through $90^{0}.$ $\phi _{1}$
and $\phi _{2}$ are the phases of waves moving along the paths $OM_{1}O$ and 
$OM_{2}O,$ respectively.\bigskip

\noindent \textbf{3.1 The Nonrelativistic Approach }

In the nonrelativistic approach the speed of light in the preferred frame is 
$c.$ Then, on the ether hypothesis, one can determine the speed of light, in
the Earth frame, i.e.,\textbf{\ }in the rest frame of the interferometer
(the $S$\ frame). This speed is $(c^{2}-v^{2})^{1/2}$ for the path along an
arm of the Michelson interferometer oriented perpendicular to its motion.
(The motion of the interferometer is at velocity $\mathbf{v}$ relative to
the preferred frame (the ether)$;$ the Earth together with the
interferometer moving with velocity $\mathbf{v}$ through the ether is
equivalent to the interferometer at rest with the ether streaming through it
with velocity $-\mathbf{v.}$) Since in $S$ both waves are brought together
to the same spatial point the phase difference $\phi _{2}-\phi _{1}$ is
determined only by the time difference $t_{2}-t_{1};$ $\phi _{2}-\phi
_{1}=2\pi (t_{2}-t_{1})/T,$ where $t_{1}$ and $t_{2}$ are the times required
for the complete trips $OM_{1}O$ and $OM_{2}O,$ respectively, and $T($=$%
\lambda /c)$ is the period of vibration of the light. From the known speed
of light one finds that $t_{1}$ is $t_{1}=t_{OM_{1}}+t_{M_{1}O}$, where $%
t_{OM_{1}}=L/c(1-v^{2}/c^{2})^{1/2}=t_{M_{1}O}$, whence 
\[
t_{1}=2L/c(1-v^{2}/c^{2})^{1/2}. 
\]
(In the following we shall denote that $t_{OM_{1}}=t_{11}$ and $%
t_{M_{1}O}=t_{12}$.) Similarly, the speed of light on the path $OM_{2}$ is $%
c-v,$ and on the return path is $c+v,$ giving that $%
t_{2}=t_{OM_{2}}+t_{M_{2}O}=t_{21}+t_{22}=L/(c-v)+L/(c+v)$. Thence 
\[
t_{2}=2L/c(1-v^{2}/c^{2}). 
\]
We see that according to the nonrelativistic approach the time $t_{1}$ is a
little less than the time $t_{2},$ even though the mirrors $M_{1}$ and $%
M_{2} $ are equidistant from $O.$ The time difference $t_{2}-t_{1}$ is 
\[
t_{2}-t_{1}=(2L/c)\gamma (\gamma -1), 
\]
($\gamma =(1-v^{2}/c^{2})^{-1/2}$) and to order $v^{2}/c^{2}$ it is $%
t_{2}-t_{1}\simeq (L/c)(v^{2}/c^{2}).$ The phase difference $\phi _{2}-\phi
_{1}$ is determined as $\phi _{2}-\phi _{1}=\omega (t_{21}+t_{22})-\omega
(t_{11}+t_{12})=\omega (t_{2}-t_{1})$ and the change in the phase difference
when the interferometer is rotated through $90^{0}$ is $\bigtriangleup (\phi
_{2}-\phi _{1})=2\omega (t_{2}-t_{1}).$ Inserting it into $\bigtriangleup N$
(\ref{delfi}) ($\omega =2\pi c/\lambda $, and the measured quantity $%
\bigtriangleup N$ is in this case $\bigtriangleup N=2(t_{2}-t_{1})c/\lambda $%
) we find that $\bigtriangleup N=(4L/\lambda )\gamma (\gamma -1).$ To the
same order $v^{2}/c^{2}$ this $\bigtriangleup N$ is 
\[
\bigtriangleup N\simeq (2L/\lambda )(v^{2}/c^{2}). 
\]
This result is obtained by the classical analysis in the Earth frame (the
interferometer rest frame).

Let us now consider the same experiment in the preferred frame (the $%
S^{\prime }$\ frame). In the nonrelativistic theory the two frames are
connected by the Galilean transformations. Consequently the corresponding
times in both frames are equal, $t_{1}=t_{1}^{\prime }$ and $%
t_{2}=t_{2}^{\prime },$ whence $t_{2}-t_{1}=t_{2}^{\prime }-t_{1}^{\prime }$
and, supposing that again the phase difference is determined only by the
time difference, $\bigtriangleup N^{\prime }=\bigtriangleup N.$ However, for
the further purposes, it is worth to find explicitly $t_{1}^{\prime }$ and $%
t_{2}^{\prime }$ considering the experiment directly in the preferred frame.
Since the speed of light in the preferred frame is $c,$ the preferred-frame
observer considers that the light travels a distance $ct_{OM_{1}^{\prime
}}^{\prime }$ along the hypotenuse of a triangle; in the same time $%
t_{OM_{1}^{\prime }}^{\prime }=t_{11}^{\prime }$ the mirror $M_{1}$ moves to 
$M_{1}^{\prime }$, i.e., to the right a distance $vt_{11}^{\prime }.$ From
the right triangle this observer finds $t_{11}^{\prime
}=L/c(1-v^{2}/c^{2})^{1/2}.$ The return trip is again along the hypotenuse
of a triangle\emph{\ }and the return time $t_{M_{1}^{\prime }O^{\prime
}}^{\prime }=t_{12}^{\prime }$ is\emph{\ }$=t_{11}^{\prime }$ (the
half-silvered mirror $O$ moved to $O^{\prime }$ in $t_{1}^{\prime }$). The
total time for such a zigzag path is, as it must be, $t_{1}^{\prime }=$ $%
t_{11}^{\prime }+t_{12}^{\prime }=t_{1}$. For the arm oriented parallel to
its motion the preferred-frame observer considers that the light, when going
from $O$ to $M_{2}^{\prime }$, must traverse a distance $L+vt_{OM_{2}^{%
\prime }}^{\prime }$ (we denote $t_{OM_{2}^{\prime }}^{\prime }$ as $%
t_{21}^{\prime }$) at the speed $c,$ whence $L+vt_{21}^{\prime
}=ct_{21}^{\prime }$ and $t_{21}^{\prime }=L/(c-v)$. The time $%
t_{M_{2}^{\prime }O^{\prime \prime }}^{\prime }$ is denoted as $%
t_{22}^{\prime }$ (the half-silvered mirror $O$ moved to $O^{\prime \prime }$
in $t_{2}^{\prime }$). Then, in a like manner, the time $t_{22}^{\prime }$
for the return trip is $t_{22}^{\prime }=L/(c+v)$. The total time $%
t_{2}^{\prime }=t_{21}^{\prime }+t_{22}^{\prime }$ is, as it must be, equal
to $t_{2},$ $t_{2}^{\prime }=t_{2}$. The phase difference in $S^{\prime }$
is determined as $\phi _{2}^{\prime }-\phi _{1}^{\prime }=\omega
(t_{21}^{\prime }+t_{22}^{\prime })-\omega (t_{11}^{\prime }+t_{12}^{\prime
})=\omega (t_{2}^{\prime }-t_{1}^{\prime }),$ and it is $\phi _{2}^{\prime
}-\phi _{1}^{\prime }=\omega (t_{2}-t_{1})$. Consequently the expected
maximum shift in the number of fringes on a $90^{0}$\ rotation in the $%
S^{\prime }$ frame is 
\begin{equation}
\bigtriangleup N^{\prime }=\bigtriangleup N\simeq (2L/\lambda )(v^{2}/c^{2}).
\label{enclas}
\end{equation}
It has to be noted here that the same $\omega ,$ i.e., $T,$ or $\lambda $,
is used for all paths, both in $S$ and $S^{\prime }.$ This discussion shows
that \emph{the nonrelativistic theory} is a consistent theory giving the
same $\bigtriangleup N$ in both frames. However it \emph{does not agree with
the experiment.} Namely Michelson and Morley found from their experiment%
\emph{\ that was no observable fringe shift.\bigskip }

\noindent \textbf{3.2 The Traditional ''AT Relativity'' Approach }

Next we examine the same experiment in the traditional ''AT relativity''
approach. We remark that the experiment is usually discussed only in the
''e'' coordinatization, and again, as in the nonrelativistic theory, the
phase difference $\phi _{2}-\phi _{1}$ is considered to be determined only
by the time difference $t_{2}-t_{1}.$ In the ''AT relativity'' and in the
''e'' coordinatization it is postulated (Einstein's second postulate), in
contrast to the nonrelativistic theory, that light \emph{always} travels
with speed $c.$

Hence in the $S$\ frame (the rest frame of the interferometer),\ and with
the same notation as in the preceding section, we find $%
t_{1}=t_{11}+t_{12}=L/c+L/c=2L/c$ and also $t_{2}=t_{21}+t_{22}=2L/c=t_{1}.$
With the assumption that only the time difference $t_{2}-t_{1}$ matters, it
follows that $\phi _{2}-\phi _{1}=\omega (t_{21}+t_{22})-\omega
(t_{11}+t_{12})=\omega (t_{2}-t_{1})=0$, whence $\bigtriangleup N=0,$ in
agreement with the experiment.

In the $S^{\prime }$\ frame (the preferred frame)\textbf{\ }the time $%
t_{1}^{\prime }$ is determined in the same way as in the nonrelativistic
theory, i.e., supposing that a zigzag path is taken by the light beam in a
moving ''light clock''. Thus, the light-travel time $t_{1}^{\prime }$ is
exactly equal to that one in the nonrelativistic theory, $t_{1}^{\prime
}=t_{11}^{\prime }+t_{12}^{\prime }=2L/c(1-v^{2}/c^{2})^{1/2}.$ Comparing
with \textbf{\ }$t_{1}=2L/c$ we see that, in contrast to the nonrelativistic
theory, it takes a longer time for light to go from end to end in the moving
clock but in the stationary clock, 
\begin{equation}
t_{1}^{\prime }=t_{1}/(1-v^{2}/c^{2})^{1/2}=\gamma t_{1}.  \label{tecr1}
\end{equation}
This relation is Eq. (\ref{tidil}) for the dilatation of time in the ''e''
coordinatization that is considered in Sec. 2.2. The presented derivation is
the usual way in which it is shown how, in the ''AT relativity,'' the time
dilatation is forced upon us by the constancy of the speed of light, see
also, e.g., Ref. 15 p.15-6 and Ref. 16 p.359, or an often cited paper on
modern tests of special relativity.$^{(17)}$ However, in the ''AT
relativity,'' the light-travel time $t_{2}^{\prime }$ is determined by
invoking the Lorentz contraction. It is argued that a preferred frame
observer measures the length of the arm oriented parallel to its motion to
be contracted to a length $L^{\prime }=L(1-v^{2}/c^{2})^{1/2}$, Eq. (\ref
{apcon}) in Sec. 2.2. Then $t_{2}^{\prime }$ is determined in the same way
as in the nonrelativistic theory but with $L^{\prime }$ replacing the rest
length $L,$ $t_{2}^{\prime }=t_{21}^{\prime }+t_{22}^{\prime }=(L^{\prime
}/(c-v))+(L^{\prime }/(c+v))$, i.e., 
\begin{equation}
t_{2}^{\prime }=2L/c(1-v^{2}/c^{2})^{1/2}=\gamma t_{2}=t_{1}^{\prime }.
\label{tcr2}
\end{equation}
Thence $t_{2}^{\prime }-t_{1}^{\prime }=0$ and 
\[
\phi _{2}^{\prime }-\phi _{1}^{\prime }=\omega (t_{21}^{\prime
}+t_{22}^{\prime })-\omega (t_{11}^{\prime }+t_{12}^{\prime })=\omega
(t_{2}^{\prime }-t_{1}^{\prime })=0, 
\]
where again, in the same way as in the nonrelativistic theory, the same $%
\omega ,$ i.e., $T,$ or $\lambda $, is used for all paths, both in $S$ and $%
S^{\prime }.$ As a consequence it is found in the ''AT relativity'' that $%
\bigtriangleup N^{\prime }$ in the $S^{\prime }$ frame is the same as $%
\bigtriangleup N$ in the $S$ frame 
\begin{equation}
\bigtriangleup N^{\prime }=\bigtriangleup N=0.  \label{traen}
\end{equation}
We quoted such usual derivation in order to illustrate how the time
dilatation and the Lorentz contraction are used in the ''AT relativity'' to
show the agreement between the theory and the famous Michelson-Morley
experiment. Although this procedure is generally accepted by the majority of
physicists as a correct one and quoted in all textbooks on the subject, we
note that such an explanation of the null result of the experiment is very
awkward and does not use at all the 4D symmetry of the spacetime. The
derivation deals with the temporal and spatial distances as well defined
quantities, i.e., in a similar way as in the prerelativistic physics, and
then in an artificial way introduces the changes in these distances due to
the motion. Our results from Secs. 2., 2.1, and 2.2 reveal why the Lorentz
contraction and the time dilatation are not physically correctly defined
transformations. Consequently the approach which uses them for the
explanation of the experimental results cannot be in agreement with the 4D
symmetry of the 4D spacetime.\bigskip

\noindent \textbf{3.3 Driscoll's ''AT Relativity'' Approach }

In Ref. 1 the above discussed ''AT relativity'' calculation in the ''e''
coordinatization (Sec. 3.2) of the fring shift in the Michelson-Morley
experiment is repeated, and, of course, the observed null fringe shift is
obtained. This result is independent of changes of $\mathbf{v}$, the
relative velocity of $S$ (the rest frame of the interferometer)\textbf{\ }%
and $S^{\prime }$ (in $S^{\prime }$ the interferometer is moving)$,$ and/or $%
\theta ,$ the angle that the undivided ray from the source to the beam
divider makes with $\mathbf{v}$. However, it is noticed$^{(1)}$ that in such
a traditional calculation of $\bigtriangleup (\phi _{2}-\phi _{1})$ only
path lengths (optical or geometrical), i.e., the temporal distances (for
example, the times $t_{1}$ and $t_{2}$ required for the complete trips $%
OM_{1}O$ and $OM_{2}O,$ respectively), are considered. The Doppler effect on
wavelength in the $S^{\prime }$ frame, in which the interferometer is
moving, is not taken into account.

Then the same calculation of $\bigtriangleup (\phi _{2}^{\prime }-\phi
_{1}^{\prime })$ as the traditional one is performed,$^{(1)}$ but determing
the increment of phase along some path, e.g. $OM_{1}^{\prime }$ in $%
S^{\prime }$, not only by the segment of geometric path length (i.e., the
temporal distance for that path) but also by the wavelength in that segment
(i.e., the frequency of the wave in that segment). Accordingly, the phase
difference (in our notation) $\phi _{1}^{\prime }-\phi _{2}^{\prime },$ in
the $S^{\prime }$ frame, between the ray along the vertical path $%
OM_{1}^{\prime }O^{\prime }$ and that one along the longitudinal path $%
OM_{2}^{\prime }O^{\prime \prime }$ respectively, is found (see$^{(1)}$) to
be 
\begin{eqnarray}
(\phi _{1}^{\prime }-\phi _{2}^{\prime })_{(b)}/2\pi  &=&2(L\nu
/c)(1+\varepsilon +\beta ^{2})-2(L\nu /c)(1+2\beta ^{2})  \nonumber \\
&=&2(L\nu /c)(\varepsilon -\beta ^{2}),  \label{drisbe}
\end{eqnarray}
Eqs.(23-25) in Ref. 1. $L$ is the length of the segment $OM_{2}$ and $%
\overline{L}=L(1+\varepsilon )$ ($\varepsilon \ll 1$) is taken in Ref. 1 to
be the length of the arm $OM_{1}.$ As explained:$^{(1)}$ ''The difference $%
\overline{L}-L=\varepsilon L$ is usually a few wavelengths ($\prec 25$) and
is essential for obtaining useful interference fringes.'' $L,$ $\overline{L}$
and $\nu $ are determined in $S$, the rest frame of the interferometer. In
this expression the Doppler effect of $\mathbf{v}$ on the frequencies, \emph{%
and the Lorentz contraction of the longitudinal arm}, are taken into
account. In a like manner Driscoll finds the phase difference in the case
when the interferometer is rotated through $90^{0}$ 
\begin{eqnarray}
(\phi _{1}^{\prime }-\phi _{2}^{\prime })_{(a)}/2\pi  &=&2(L\nu
/c)(1+\varepsilon +2\beta ^{2})-2(L\nu /c)(1+\beta ^{2})  \nonumber \\
&=&2(L\nu /c)(\varepsilon +\beta ^{2}),  \label{drisca}
\end{eqnarray}
Eqs.(19-21) in Ref. 1. Hence it is found$^{(1)}$ a ''surprising'' non-null
fringe shift 
\begin{equation}
\bigtriangleup N^{\prime }=\bigtriangleup (\phi _{2}^{\prime }-\phi
_{1}^{\prime })/2\pi =4(L\nu /c)\beta ^{2},  \label{shift}
\end{equation}
where $\bigtriangleup (\phi _{2}^{\prime }-\phi _{1}^{\prime })=(\phi
_{1}^{\prime }-\phi _{2}^{\prime })_{(b)}-(\phi _{1}^{\prime }-\phi
_{2}^{\prime })_{(a)},$ and we see that the entire fringe shift is due to
the Doppler shift. From the non-null result (\ref{shift}) the author of$%
^{(1)}$ concluded: ''that the Maxwell-Einstein electromagnetic equations and
special relativity jointly are disproved, not confirmed, by the
Michelson-Morley experiment.'' However such a conclusion cannot be drawn
from the result (\ref{shift}). The origin of the appearance of $%
\bigtriangleup N^{\prime }\neq 0$ (\ref{shift}) is quite different than that
considered in Ref. 1, and it will be explained below. While$^{(1)}$
investigates those changes which are caused by the Doppler effect another
work$^{(18)}$ considers the changes in the usual derivation of $%
\bigtriangleup N^{\prime },$ which are caused by the aberration of light.
Both changes are examined only in the ''e'' coordinatization, and both would
be different in, e.g., the ''r'' coordinatization. This means that $%
\bigtriangleup N^{\prime }$ in $S^{\prime }$ will be dependent on the chosen
synchronization. Also, both works$^{(1,18)}$ deal with the Lorentz
contraction in the same way as in the traditional analysis. But the Lorentz
contraction is an AT, as shown in Secs. 2., 2.1, and 2.2. Consequently, the
traditional analysis and the works$^{(1,18)}$ belong to the ''AT
relativity,'' which, as found in Ref. 8, and as follows from the dependence
of the theoretical results on the chosen synchronization, is not capable to
explain in a satisfactory manner the results of the Michelson-Morley
experiment.\bigskip \medskip 

\noindent {\large 4. THE ''TT RELATIVITY'' APPROACH }

Next we examine the Michelson-Morley experiment from the ''TT relativity''
viewpoint. The relevant quantity is the phase of a light wave, and it is
(when written in the abstract index notation) 
\begin{equation}
\phi =k^{a}g_{ab}l^{b},  \label{phase}
\end{equation}
where $k^{a}$ is the propagation 4-vector, $g_{ab}$ is the metric tensor and 
$l^{b}$ is the distance 4-vector. All quantities in (\ref{phase}) are true
tensor quantities. As discussed in Sec. 2. these quantities can be written
in the coordinate-based geometric language and, e.g., the decompositions of $%
k^{a}$ in $S$ and $S^{\prime }$ and in the ''e'' and ''r'' coordinatizations
are 
\[
k^{a}=k_{e}^{\mu }e_{\mu }=k_{e}^{\mu ^{\prime }}e_{\mu ^{\prime
}}=k_{r}^{\mu }r_{\mu }=k_{r}^{\mu ^{\prime }}r_{\mu ^{\prime }},
\]
where the basis components $k^{\mu }$ of the CBGQ in the ''e''
coordinatization are transformed by $L^{\mu ^{\prime }}{}_{\nu ,e}$ (\ref
{lorus}), while the basis vectors $e_{\mu }$ are transformed by the inverse
transformation $(L^{\mu ^{\prime }}{}_{\nu ,e})^{-1}=L^{\mu }{}_{\nu
^{\prime },e}.$ Similarly holds for the ''r'' coordinatization where the
Lorentz transformation $L^{\mu ^{\prime }}{}_{\nu ,r}$ (\ref{elr}) has to be
used. By the same reasoning the phase $\phi $ (\ref{phase}) is given in the
coordinate-based geometric language as 
\begin{equation}
\phi =k_{e}^{\mu }g_{\mu \nu ,e}\,l_{e}^{\nu }=k_{e}^{\mu ^{\prime }}g_{\mu
\nu ,e}\,l_{e}^{\nu ^{\prime }}=k_{r}^{\mu }g_{\mu \nu ,r}\,l_{r}^{\nu
}=k_{r}^{\mu ^{\prime }}g_{\mu \nu ,r}\,l_{r}^{\nu ^{\prime }},  \label{pha2}
\end{equation}
(Note that the Lorentz transformation $L^{\mu ^{\prime }}{}_{\nu ,e}$ (\ref
{lorus}) and also $L^{\mu ^{\prime }}{}_{\nu ,r}$ (\ref{elr}) are the TT,
i.e., the isometries, and hence $g_{\mu \nu ,e}=g_{\mu ^{\prime }\nu
^{\prime },e}$, $g_{\mu \nu ,r}=g_{\mu ^{\prime }\nu ^{\prime },r}$, what is
already taken into account in (\ref{pha2}).) The traditional derivation of $%
\bigtriangleup N$ (Sec. 3.2) deals, as already said, only with the
calculation of $t_{1}$ and $t_{2}$ in $S$ and $t_{1}^{\prime }$ and $%
t_{2}^{\prime }$ in $S^{\prime },$ but does not take into account either the
changes in frequencies due to the Doppler effect or the aberration of light.
The ''AT relativity'' calculations$^{(1,18)}$ improve the traditional
procedure taking into account the changes in frequencies,$^{(1)}$ and the
aberration of light.$^{(18)}$ But all these approaches explain the
experiments using the AT, the Lorentz contraction and the time dilatation,
and furthermore they always work only in the ''e'' coordinatization. None of
the ''AT relativity'' calculations deals with the true tensors or with the
CBGQs (comprising both components and a basis). In this case such 4D tensor
quantity is the phase (\ref{phase}) or (\ref{pha2}). It will be shown here
that the non-null theoretical result obtained in Ref. 1 is a consequence of
the fact that Driscoll's calculation also belongs to the ''AT relativity''
approach. It considers only a part of the 4D tensor quantity $\phi $ (\ref
{phase}), or (\ref{pha2}), uses the AT and works only in the ''e''
coordinatization. \emph{In the ''TT relativity'' approach to special
relativity neither the Doppler effect nor the aberration of light exist
separately as well defined physical phenomena.} \emph{The separate
contributions to }$\phi $\emph{\ (\ref{phase}), or (\ref{pha2}), of the }$%
\omega t$\emph{\ (i.e., }$k^{0}l_{0}$\emph{)} \emph{factor}$^{(1)}$\emph{\
and }$\mathbf{kl}$\emph{\ (i.e., }$k^{i}l_{i}$) \emph{factor}$^{(18)}$\emph{%
\ are, in general case, meaningless in the ''TT relativity.'' From the ''TT
relativity'' viewpoint only their indivisible unity, the phase }$\phi $\ 
\emph{(\ref{phase}), or (\ref{pha2}), is a correctly defined 4D quantity. }%
All quantities in (\ref{phase}), i.e., $k^{a}$, $g_{ab},$ $l^{b}$ and $\phi ,
$ are the true tensor quantities, which means that in all relatively moving
IFRs and in all permissible coordinatizations always \emph{the same 4D
quantity}, e.g., $k^{a},$ or $l^{b},$ or $\phi ,$\ is considered. (Eq. (\ref
{pha2}) shows it for $\phi $.) This is not the case in the ''AT relativity''
where, for example, the relation $t_{1}^{\prime }=\gamma t_{1}$ is not the
Lorentz transformation of some 4D quantity, and $t_{1}^{\prime }$ and $t_{1}$
do not correspond to the same 4D quantity considered in $S^{\prime }$ and $S$
respectively but to different 4D quantities, as can be clearly seen from
Sec. 2.2 (see Eq. (11)). Only in the ''e'' coordinatization the $\omega t$
and $\mathbf{kl}$ factors can be considered separately. Therefore, and in
order to retain the similarity with the prerelativistic and the ''AT
relativity'' considerations, we first determine $\phi $ (\ref{phase}), (\ref
{pha2}), in the ''e'' coordinatization and in the $S$ frame (the rest frame
of the interferometer). This means that $\phi $ will be calculated from (\ref
{pha2}) as the CBGQ $\phi =k_{e}^{\mu }g_{\mu \nu ,e}\,l_{e}^{\nu }.$

Let now $A,$ $B$ and $A_{1}$ denote the events; the departure of the
transverse ray from the half-silvered mirror $O,$ the reflection of this ray
on the mirror $M_{1}$ and the arrival of this beam of light after the round
trip on the half-silvered mirror $O,$ respectively. In the same way we have,
for the longitudinal arm of the inteferometer, the corresponding events $A,$ 
$C$ and $A_{2}.$ To simplify the notation we omit the subscript 'e' in all
quantities. Then $k_{AB}^{\mu }$ and $l_{AB}^{\mu }$ (the basis components
of $k_{AB}^{a}$ and $l_{AB}^{a}$ in the ''e'' coordinatization and in $S$)
for the wave on the trip $OM_{1}$ (the events $A$ and $B$) are $k_{AB}^{\mu
}=(\omega /c,0,2\pi /\lambda ,0),$ $l_{AB}^{\mu }=(ct_{M_{1}},0,\overline{L}%
,0)$. For the wave on the return trip $M_{1}O,$ (the events $B$ and $A_{1}$) 
$k_{BA_{1}}^{\mu }=(\omega /c,0,-2\pi /\lambda ,0)$ and $l_{BA_{1}}^{\mu
}=(ct_{M_{1}},0,-\overline{L},0)$. Hence the increment of phase $\phi _{1}$
for the the round trip $OM_{1}O,$ is 
\begin{equation}
\phi _{1}=k_{AB}^{\mu }\,l_{\mu AB}+k_{BA_{1}}^{\mu }l_{\mu
BA_{1}}=2(-\omega t_{M_{1}}+(2\pi /\lambda )\overline{L}),  \label{fijd}
\end{equation}
where $\omega $ is the angular frequency and, for the sake of comparison
with,$^{(1)}$ the length of the arm $OM_{1}$ is taken to be $\overline{L}%
=L(1+\varepsilon ),$ and $L$ is the length of the segment $OM_{2}.$ Using
the Lorentz transformation $L^{\mu ^{\prime }}{}_{\nu ,e}$ (\ref{lorus}) one
can find $k^{\mu ^{\prime }}$ and $l^{\mu ^{\prime }}$ in the ''e''
coordinatization and in $S^{\prime }$ for the same trips as in $S$. Then it
can be easily shown that $\phi _{1}^{\prime }$ in $S^{\prime }$ is the same
as in $S,$ $\phi _{1}^{\prime }=\phi _{1}.$ Also using the transformation
matrix $T_{\;\nu ,r}^{\mu }$ (Sec. 2), which transforms the ''e''
coordinatization to the ''r'' coordinatization, one can get all quantities
in the ''r'' coordinatization and in $S$, and then by the Lorentz
transformation $L^{\mu ^{\prime }}{}_{\nu ,r}$ (\ref{elr}) these quantities
can be determined in the ''r'' coordinatization and in $S^{\prime }$. $\phi
_{1}$ will be always the same in accordance with (\ref{pha2}). Note that $%
g_{\mu \nu ,r}$ from Sec. 2 has to be used in the calculation of $\phi $ in
the ''r'' coordinatization. As an example we quote $k_{AB,r}^{\mu }$ and $%
l_{AB,r}^{\mu }$: $k_{AB,r}^{\mu }=((\omega /c)-2\pi /\lambda ,0,2\pi
/\lambda ,0)$ and $l_{AB,r}^{\mu }=(ct_{M_{1}}-\overline{L},0,\overline{L}%
,0).$ Hence, using $g_{\mu \nu ,r}$ one easily finds that 
\[
\phi _{AB,r}=k_{r}^{\mu }g_{\mu \nu ,r}\,l_{r}^{\nu }=(-\omega
t_{M_{1}}+(2\pi /\lambda )\overline{L})=\phi _{AB,e}.
\]
For further purposes we shall also need $k_{AB,r}^{\mu ^{\prime }}$ and $%
l_{AB,r}^{\mu ^{\prime }}.$ They are $k_{AB,r}^{\mu ^{\prime }}=((\gamma
\omega /c)(1+\beta )-2\pi /\lambda ,-\beta \gamma \omega /c,2\pi /\lambda ,0)
$ and $l_{AB,r}^{\mu ^{\prime }}=(\gamma ct_{M_{1}}(1+\beta )-\overline{L}%
,-\beta \gamma ct_{M_{1}},\overline{L},0)$ which yields 
\[
\phi _{AB,r}^{\prime }=\phi _{AB,r}=\phi _{AB,e}^{\prime }=\phi _{AB,e}.
\]
In a like manner we find $k_{AC}^{\mu }$ and $l_{AC}^{\mu }$ for the wave on
the trip $OM_{2},$ (the corresponding events are $A$ and $C$) as $%
k_{AC}^{\mu }=(\omega /c,2\pi /\lambda ,0,0)$ and $l_{AC}^{\mu
}=(ct_{M_{2}},L,0,0).$ For the wave on the return trip $M_{2}O$ (the
corresponding events are $C$ and $A_{2}$) $k_{CA_{2}}^{\mu }=(\omega
/c,-2\pi /\lambda ,0,0)$ and $l_{CA_{2}}^{\mu }=(ct_{M_{2}},-L,0,0)$),
whence 
\begin{equation}
\phi _{2}=k_{AC}^{\mu }\,l_{\mu AC}+k_{CA_{2}}^{\mu }l_{\mu
CA_{2}}=2(-\omega t_{M_{2}}+(2\pi /\lambda )L).  \label{f2jd}
\end{equation}
Of course one finds the same $\phi _{2}$ in $S$ and $S^{\prime }$ and in the
''e'' and ''r'' coordinatizations. Hence 
\begin{equation}
\phi _{1}-\phi _{2}=-2\omega (t_{M_{1}}-t_{M_{2}})+2(2\pi /\lambda )(%
\overline{L}-L).  \label{phasdif}
\end{equation}
Particularly for $\overline{L}=L,$ and consequently $t_{M_{1}}=t_{M_{2}},$
one finds $\phi _{1}-\phi _{2}=0.$ It can be easily shown that the same
difference of phase (\ref{phasdif}) is obtained in the case when the
interferometer is rotated through $90^{0},$ whence we find that $%
\bigtriangleup (\phi _{1}-\phi _{2})=0,$ and $\bigtriangleup N=0.$ \emph{%
According to the construction }$\phi $\emph{\ (\ref{phase}), or (\ref{pha2}%
), is a frame independent quantity and it also does not depend on the chosen
coordinatization in a considered IFR.} Thus we conclude that 
\begin{equation}
\bigtriangleup N_{e}=\bigtriangleup N_{e}^{\prime }=\bigtriangleup
N_{r}=\bigtriangleup N_{r}^{\prime }=0.  \label{delshif}
\end{equation}
This result is in a complete agreement with the Michelson-Morley$^{(2)}$
experiment.\bigskip 

\noindent \textbf{4.1 Explanation of Driscoll's Non-Null Fringe Shift }

Driscoll's improvement of the traditional ''AT relativity'' derivation of
the fringe shift \emph{can be easily obtained from our ''TT relativity''
approach taking only the product }$k_{e}^{0^{\prime }}l_{0^{\prime }e}$\emph{%
\ in the calculation of the increment of phase }$\phi _{e}^{\prime }$\emph{\
in} $S^{\prime }$ in which the apparatus is moving. All quantities in the
''e'' coordinatization and in $S^{\prime }$ are obtained by the Lorentz
transformation $L^{\mu ^{\prime }}{}_{\nu ,e}$ (\ref{lorus}) from the
corresponding ones in $S.$ We remark once again that Driscoll's ''AT
relativity'' approach refers only to the ''e'' coordinatization (of course
it holds for the traditional ''AT relativity'' approach from Sec. 3.2 as
well). Therefore we again omit the subscript $^{\prime }e^{\prime }$ in all
quantities. Then we find that in the $S^{\prime }$ frame $k_{AB}^{\mu
^{\prime }}=(\gamma \omega /c,-\beta \gamma \omega /c,2\pi /\lambda ,0)$ and 
$l_{AB}^{\mu ^{\prime }}=(\gamma ct_{M_{1}},-\beta \gamma ct_{M_{1}},%
\overline{L},0),$ and also $k_{BA_{1}}^{\mu ^{\prime }}=(\gamma \omega
/c,-\beta \gamma \omega /c,-2\pi /\lambda ,0)$ and $l_{BA_{1}}^{\mu ^{\prime
}}=(\gamma ct_{M_{1}},-\beta \gamma ct_{M_{1}},-\overline{L},0),$ giving
that 
\begin{equation}
(-1/2\pi )(k_{AB}^{0^{\prime }}l_{0^{\prime }AB}+k_{BA_{1}}^{0^{\prime
}}l_{0^{\prime }BA_{1}})=2\gamma ^{2}\nu t_{M_{1}}\simeq 2(L\nu
/c)(1+\varepsilon +\beta ^{2}),  \label{truedri}
\end{equation}
which is exactly Driscoll's result $\bigtriangleup P_{Hb}$; for our notation
see (\ref{drisbe}). Similarly one finds that 
\begin{eqnarray}
(-1/2\pi )(k_{AC}^{0^{\prime }}l_{0^{\prime }AC}+k_{CA_{2}}^{0^{\prime
}}l_{0^{\prime }CA_{2}}) &=&2\gamma ^{2}(\nu t_{M_{2}}+\beta ^{2}L/\lambda )
\nonumber \\
&\simeq &2(L\nu /c)(1+2\beta ^{2}),  \label{tr2}
\end{eqnarray}
which is Driscoll's result $\bigtriangleup P_{\Xi b},$ see (\ref{drisbe}).
In the same way we can find in $S^{\prime }$ Driscoll's result (\ref{drisca}%
) and finally the non-null fringe shift (\ref{shift}).

We remark that the non-null fringe shift (\ref{shift}) would be quite
different in another coordinatization, e.g., in the ''r'' coordinatization,
since only a part $k_{e}^{0^{\prime }}l_{0^{\prime }e}$ of the whole 4D
tensor quantity $\phi $ (\ref{phase}) or (\ref{pha2}) is considered. The
basis components of the metric tensor in the ''r'' coordinatization, i.e., $%
g_{\mu \nu ,r},$ do not form a diagonal matrix and therefore the temporal
and spatial parts of $\phi =k_{r}^{\mu }g_{\mu \nu ,r}\,l_{r}^{\nu }$ cannot
be separated. From the above expressions we can easily show that \emph{the
part} $k_{r}^{0^{\prime }}g_{00,r}\,l_{r}^{0^{\prime }}$ (i.e., $%
k_{r}^{0^{\prime }}l_{0^{\prime }\,r}$) \emph{is quite different but the
Driscoll's expression} $k_{e}^{0^{\prime }}g_{00,e}\,l_{e}^{0^{\prime }}$
(i.e., $k_{e}^{0^{\prime }}l_{0^{\prime }\,e}$). However the physics must
not depend on the chosen coordinatization. Thus when only a part of the
whole phase $\phi $ (\ref{phase}) or (\ref{pha2}) is taken into account then
it leads to an unphysical result.

\emph{The same calculation of} $k^{i^{\prime }}l_{i^{\prime }},$ \emph{the
contribution of the spatial parts of} $k^{\mu ^{\prime }}$ \emph{and} $%
l_{\mu ^{\prime }}$ \emph{to} $\bigtriangleup N_{e}^{\prime },$ \emph{shows
that this term exactly cancel the} $k^{0^{\prime }}l_{0^{\prime }}$ \emph{%
contribution (Driscoll's non-null fringe shift (\ref{shift})), yielding that}
$\bigtriangleup N_{e}^{\prime }=\bigtriangleup N_{e}=0.$ Thus the ''TT
relativity''approach to SR naturally explaines the reason for the existence
of Driscoll's non-null fringe shift (\ref{shift}).\bigskip

\noindent \textbf{4.2 Explanation of the ''Apparent'' Agreement Between the
Traditional Analysis and the Experiment }

The results of the usual ''AT relativity'' calculation, which are presented
in Sec. 3.2, \emph{can be easily explained from our true tensor formulation
of SR taking only the part }$k_{e}^{0}l_{0^{\prime }e}$\ \emph{of the whole
phase }$\phi $ \emph{(\ref{phase}) or (\ref{pha2}) in the calculation of the
increment of phase }$\phi _{e}^{\prime }$\emph{\ in} $S^{\prime }.$ In
contrast to Driscoll's treatment the traditional analysis considers the part 
$k_{e}^{0}l_{0e}$ (of the whole phase $\phi $ (\ref{phase}), (\ref{pha2}))
in $S,$ the rest frame of the interferometer, and $k_{e}^{0}l_{0^{\prime }e}$%
\ in $S^{\prime }$, in which the apparatus is moving. $k_{e}^{0}$\emph{\ is
not changed in transition from} $S$\emph{\ to} $S^{\prime }$. Thus the
increment of phase $\phi _{1}$ for the round trip $OM_{1}O$ in $S$, is 
\begin{equation}
\phi
_{1}=k_{AB}^{0}\,g_{00,e}l_{AB}^{0}+k_{BA_{1}}^{0}g_{00,e}l_{BA_{1}}^{0}=-2(%
\omega /c)(ct_{M_{1}})=-2\omega t_{M_{1}}.  \label{apfi1}
\end{equation}
In the $S^{\prime }$ frame we find for the same trip that 
\begin{equation}
\phi _{1}^{\prime }=k_{AB}^{0}\,l_{0^{\prime }AB}+k_{BA_{1}}^{0}l_{0^{\prime
}BA_{1}}=-2(\omega /c)(\gamma ct_{M_{1}})=-2\omega (\gamma t_{M_{1}}).
\label{apf1cr}
\end{equation}
This is exactly the result obtained in the traditional analysis in Sec. 3.2,
which is inerpreted as that there is a ''time dilatation'' $t_{1}^{\prime
}=\gamma t_{1}$. In the same way we find that the increment of phase $\phi
_{2}$ for the round trip $OM_{2}O$ in $S$, is 
\begin{equation}
\phi _{2}=k_{AC}^{0}\,l_{0AC}+k_{CA_{2}}^{0}l_{0CA_{2}}=-2\omega t_{M_{2}},
\label{apfi2}
\end{equation}
and $\phi _{2}^{\prime }$ in $S^{\prime }$ is 
\begin{equation}
\phi _{2}^{\prime }=k_{AC}^{0}\,l_{0^{\prime }AC}+k_{CA_{2}}^{0}l_{0^{\prime
}CA_{2}}=-2(\omega /c)(\gamma ct_{M_{2}})=-2\omega (\gamma t_{M_{2}}).
\label{af2cr}
\end{equation}
This is again the result of the traditional analysis, the ''time
dilatation,'' $t_{2}^{\prime }=\gamma t_{2}$. For $t_{1}=t_{2}$, i.e., for $%
\overline{L}=L,$ one finally finds the null fringe shift that is obtained in
the traditional analysis $\bigtriangleup N_{e}^{\prime }=\bigtriangleup
N_{e}=0.$ We see that \emph{such a null fringe shift is obtained taking into
account only a part}\ \emph{of the whole phase }$\phi $ \emph{(\ref{phase})
or (\ref{pha2}), and additionally, in that part, }$k_{e}^{0}$\emph{\ is not
changed in transition from} $S$\emph{\ to} $S^{\prime }$. Obviously this
correct result follows from a physically incorrect treatment the phase $\phi 
$ (\ref{phase}) or (\ref{pha2}). Furthermore it has to be noted that the
usual calculation is always done only in the ''e'' coordinatization.

Since only the part $k_{e}^{0}l_{0e}$\ of the whole phase $\phi $ (\ref
{phase}) or (\ref{pha2}) is taken into account (and also $k_{e}^{0^{\prime
}}=k_{e}^{0}$) the results of the usual ''AT relativity'' calculation are
coordinatization dependent. We explicitly show it using the ''r''
coordinatization.

In the ''r'' coordinatization the increment of phase $\phi _{r}$ is
calculated from $\phi _{r}=k_{r}^{0}g_{00,r}\,l_{r}^{0}$ in $S$ and from $%
\phi _{r}^{\prime }=k_{r}^{0}g_{00,r}\,l_{r}^{0^{\prime }}$ in $S^{\prime }.$
Hence we find that $\phi _{1r}$ for the round trip $OM_{1}O$ in $S$ is 
\begin{equation}
\phi _{1r}=-2(\omega t_{M_{1}}+(2\pi /\lambda )\overline{L}),  \label{f1er}
\end{equation}
and $\phi _{2r}$ for the round trip $OM_{2}O$ in $S$ is 
\begin{equation}
\phi _{2r}=-2(\omega t_{M_{2}}+(2\pi /\lambda )L).  \label{f2er}
\end{equation}
For $\overline{L}=L,$ and consequently $t_{M_{1}}=t_{M_{2}},$ we find that $%
\phi _{1r}-\phi _{2r}=0$, whence $\bigtriangleup N_{r}=0.$ Remark that the
phases $\phi _{1r}$ and $\phi _{2r}$ differ from the corresponding phases $%
\phi _{1e}$ and $\phi _{2e}$ in the ''e'' coordinatization. As shown above
this is not the case when the whole phase $\phi $ (\ref{phase}) or (\ref
{pha2}) is taken into account.

However, in $S^{\prime },$ we find for the same trips that 
\begin{equation}
\phi _{1r}^{\prime }=-2(\gamma \omega t_{M_{1}}(1+\beta )+(2\pi /\lambda )%
\overline{L}),  \label{efcr1}
\end{equation}
\begin{equation}
\phi _{2r}^{\prime }=-2\gamma ^{2}(1+\beta ^{2})(\omega t_{M_{2}}+(2\pi
/\lambda )L).  \label{fcrt2}
\end{equation}
Obviously $\phi _{1r}^{\prime }-\phi _{2r}^{\prime }\neq 0$ and consequently
it leads to \emph{the non-null fringe shift} 
\begin{equation}
\bigtriangleup N_{r}^{\prime }\neq 0,  \label{denrc}
\end{equation}
which holds even in the case when $t_{M_{1}}=t_{M_{2}}.$ This result clearly
shows that the agreement between the usual ''AT relativity'' calculation and
the Michelson-Morley experiment is only an ''apparent'' agreement. It is
achieved by an incorrect procedure and it holds only in the ''e''
coordinatization. We also remark that the traditional analysis, i.e., the
''AT relativity,'' gives different values for the phases, e.g., $\phi _{1e},$
$\phi _{1e}^{\prime },$ $\phi _{1r}$ and $\phi _{1r}^{\prime },$ since only
a part of the whole phase $\phi $ (\ref{phase}) or (\ref{pha2}) is
considered. \emph{These phases are frame and coordinatization dependent
quantities. When the whole phase }$\phi $\emph{\ (\ref{phase}) or (\ref{pha2}%
) is taken into account, i.e., in ''TT relativity,'' all the mentioned
phases are exactly equal quantities; they are the same,} \emph{frame and
coordinatization independent, quantity. \bigskip \bigskip \bigskip }

\noindent {\large 5. CONCLUSIONS }

In$^{(1)}$ the usual ''relativistic'' calculation of the fringe shift in the
Michelson-Morley experiment is objected on the grounds that it does not take
into account the changes in frequencies due to the Doppler effect. Our
discussion shows that Driscoll's calculation is not free from ambiguities
either. It also does not work with the complete expression for the phase of
the light waves ((\ref{phase}) or (\ref{pha2})) travelling along the arms of
the Michelson-Morley interferometer. Both calculations are shown to belong
to the ''AT relativity,'' which does not deal with the whole 4D tensor
quantities and their true transformations. In this paper we have exposed the
approach to SR that deals with true tensors and the true tensor equations
(when no basis is chosen) or equivalently with the CBGQs and equations (when
the coordinate basis is introduced), i.e., the ''TT relativity.'' This
approach uses the whole phase $\phi $ (\ref{phase}) or (\ref{pha2}) and
yields in all IFRs and in all permissible coordinatizations the observed
null fringe shift. At the same time it successfully explains an apparent
agreement (it holds only in the ''e'' coordinatization) of the traditional
''AT relativity'' approach and disagreement of Driscoll's ''AT relativity''
approach with the experimental results. They are simply consequences of the
use of only some parts of the 4D tensor quantity $\phi $ (\ref{phase}) or (%
\ref{pha2}) and the use of the AT, the Lorentz contraction and the time
dilatation, in the calculation of the increment of phase. The results of the
traditional analysis are exactly obtained taking into account only the part $%
k_{e}^{0}l_{0^{\prime }e}$\ of the whole phase $\phi $ (\ref{phase}) or (\ref
{pha2}) in the calculation of the increment of phase $\phi _{e}^{\prime }$\
in $S^{\prime }.$ Similarly the results of Driscoll's analysis are obtained
taking only the part $k_{e}^{0^{\prime }}l_{0^{\prime }e}$ in the
calculation of $\phi _{e}^{\prime }$\ in $S^{\prime }.$ In conclusion, the
analysis performed in this paper reveals that the Michelson-Morley
experiment does not confirm either the validity of the traditional Einstein
approach or the validity of Driscoll's approach. In other words, the
experiment does not confirm the ''AT relativity'' approach, but rather an
invariant ''TT relavitity'' approach to SR. \pagebreak 

\noindent \textbf{References}

\noindent \ 1. R.B. Driscoll, Phys. Essays \textbf{10}, 394 (1997).

\noindent 2. A.A. Michelson and E.W. Morley, Philos. Mag. \textbf{24}, 449
(1887); Am. J. Sci. \textbf{34}, 333 (1887).

\noindent \ 3. A. Einstein, Ann. Physik \textbf{17,} 891 (1905), tr. by W.
Perrett and G.B.

\noindent \quad \ Jeffery, in \textit{The principle of relativity} (Dover,
New York).

\noindent \ 4. F. Rohrlich, Nuovo Cimento B \textbf{45}, 76 (1966).

\noindent \ 5. T.Ivezi\'{c}, Found. Phys. Lett. \textbf{12}, 105 (1999).

\noindent \ 6. T.Ivezi\'{c}, Found. Phys. Lett. \textbf{12}, 507 (1999).

\noindent \ 7. T.Ivezi\'{c}, preprint Lanl Archives: physics/0012048; to be
published

\noindent \qquad in Found. Phys.

\noindent \ 8. T. Ivezi\'{c}, preprint Lanl Archives: physics/0007031.

\noindent \ 9. A. Gamba, Am. J. Phys. \textbf{35}, 83 (1967).

\noindent 10. R.M. Wald, \textit{General relativity }(The University of
Chicago Press, Chicago,\ 

\noindent \quad \ 1984).

\noindent 11. B.F. Schutz, \textit{A first course in general relativity}
(Cambridge University

\noindent \qquad Press, Cambridge, 1985).

\noindent 12. C.W. Misner, K.S. Thorne and J.A. Wheeler, \textit{Gravitation}%
, (Freeman,

\noindent \qquad San Francisco, 1970).

\noindent 13. C. Leubner, K. Aufinger and P. Krumm, \textit{Eur. J. Phys.} 
\textbf{13,} 170 (1992).

\noindent 14. D.E. Fahnline, Am. J. Phys. \textbf{50}, 818 (1982).

\noindent 15. R.P. Feynman, R.B. Leightonn and M. Sands, \textit{The Feynman
lectures on}

\noindent \quad \ \textit{physics,} Vol.1 (Addison-Wesley, Reading, 1964).

\noindent 16. C. Kittel, W.D. Knight and M.A. Ruderman, \textit{Mechanics}
(McGraw-Hill,

\noindent \quad \ New York, 1965).

\noindent 17. M.P. Haugan and C.M. Will, Phys. Today \textbf{40, }69 (1987).

\noindent 18. R.A. Schumacher, Am.J. Phys. \textbf{62}, 609 (1994).

\end{document}